\documentclass[11pt]{article}
\usepackage{blois,amsmath, amssymb,xcolor,epsfig}

\def\mathswitch#1{\relax\ifmmode#1\else$#1$\fi}
\def\mathswitchr#1{\relax\ifmmode{\mathrm{#1}}\else$\mathrm{#1}$\fi}
\def\mathswitchit#1{\relax\ifmmode{#1}\else$#1$\fi}

\newcommand{\mr}{\mathrm}

\newcommand{\MSbar}{\mathswitch {\overline{\mr{MS}}}}

\def\citere#1{\mbox{Ref.~\cite{#1}}}
\def\citeres#1{\mbox{Refs.~\cite{#1}}}

\def\mathswitch#1{\relax\ifmmode#1\else$#1$\fi}
\def\mathswitchr#1{\relax\ifmmode{\mathrm{#1}}\else$\mathrm{#1}$\fi}
\def\mathswitchit#1{\relax\ifmmode{#1}\else$#1$\fi}

\newcommand{\Ph}{\mathswitchr h}

\newcommand{\PW}{\mathswitchr W}
\newcommand{\PZ}{\mathswitchr Z}

\newcommand{\Prophecy}{{\sc Prophecy4f}}

\begin{document}
\vspace*{4cm}
\title{Higgs-boson decay to four fermions  in the Two-Higgs-Doublet Model \\ and \Prophecy}

\author{Talk presented at the International Workshop on Future Linear Colliders (LCWS2017), Strasbourg, France, 23-27 October 2017. C17-10-23.2.
\\[0.5cm]
L. Altenkamp$^A$, S. Dittmaier$^A$, H. Rzehak$^B$\footnote{Speaker.}}

\address{$^A$ Albert-Ludwigs-Universit\"at Freiburg, Physikalisches Institut,
  Freiburg, Germany\\
  $^B$ CP3-Origins, University of Southern Denmark, Odense, Denmark 
}

\maketitle\abstracts{We present the next-to-leading-order calculation of the partial decay widths of 
the light CP-even Higgs boson decaying into four fermions in the Two-Higgs-Doublet Model. The four different renormalization schemes applied in the calculation are described as well as the calculation and its implementation into the analysis tool \Prophecy. Some sample results show the size of the next-to-leading-order correction as well as the overall size of the deviation from the Standard Model prediction. 
}

\section{Introduction}

The Higgs-boson decay to four fermions, $\Ph\to\PW\PW/\PZ\PZ\to4f$,
is one of the best studied Higgs decay channels, in particular the decay into four charged leptons, 
which delivers a very clean experimental signal 
and plays an important role in the Higgs mass measurement \cite{Aad:2015zhl}. Improving the accuracy of the measurements requires, at the same time, improving theoretical predictions  not only of the Standard Model 
(SM), but also of possible extensions of the SM to the same level of precision. Hence higher-order corrections need to be taken into account. We present the calculation of the Higgs-boson decay to four fermions in the Two-Higgs-Doublet Model (THDM) at next-to-leading order (NLO) including electroweak (EW) as well as QCD corrections.

The THDM is one of the simplest extensions of the SM, containing also a charged Higgs boson. Many more complex models contain a second Higgs doublet. In these cases, a general THDM can be a low-energy effective theory for these models. In the following, we will restrict the calculation to a THDM with the specific assumptions described below.

\section{The Two-Higgs-Doublet Model and its renormalization}

The Higgs potential $V$ of the THDM is assumed to be
\begin{align}
V={}&m^2_{11} \Phi^{\dagger}_1 \Phi_1+ m^2_{22} \Phi^\dagger_2 \Phi_2
-m^2_{12} (\Phi^\dagger_1 \Phi_2 + \Phi^\dagger_2 \Phi_1)\nonumber\\ 
&+ \frac{1}{2} \lambda_1 (\Phi^\dagger_1 \Phi_1)^2+\frac{1}{2} \lambda_2 (\Phi^\dagger_2 \Phi_2)^2+ \lambda_3 (\Phi^\dagger_1 \Phi_1)(\Phi^\dagger_2 \Phi_2)+ \lambda_4 (\Phi^\dagger_1 \Phi_2)(\Phi^\dagger_2 \Phi_1)\nonumber\\ 
&+\frac{1}{2} \lambda_5\left[(\Phi^\dagger_1 \Phi_2)^2+(\Phi^\dagger_2 \Phi_1)^2\right], 
\label{eq:lambdapara}
\end{align}
where $\Phi_1$, $\Phi_2$ denote the two Higgs doublets, $m^2_{11}$, $m^2_{12}$, $m^2_{22}$ the mass parameters, and $\lambda_1, \dots, \lambda_5$ the quartic Higgs couplings. The symmetry of the Higgs potential under the transformation $\Phi_1 \rightarrow -\Phi_1$ is only softly broken by non-vanishing values of $m^2_{12}$ \cite{HHGuide:1990,Gunion:2005ja}. In addition, we assume CP-conservation so that all parameters in the Higgs potential are real.
The two Higgs doublets can be decomposed as
\begin{align}
\Phi_1=\begin{pmatrix} \phi_1^+ \\ \frac{1}{\sqrt{2}}(\eta_1+i \chi_1+v_1)\end{pmatrix}, && \Phi_2=\begin{pmatrix}\phi_2^+ \\ \frac{1}{\sqrt{2}}(\eta_2+i \chi_2+v_2)\end{pmatrix},
\label{eq:decom}
\end{align}
where $v_1$ and $v_2$ are the Higgs vacuum expectation values and $\phi_1^+,\phi_2^+$, $\eta_1,\eta_2$,  $\chi_1,\chi_2$ the charged, the neutral CP-even, and the neutral CP-odd fields, respectively. The fields with the same quantum numbers can mix, and the resulting mass eigenstates correspond to two CP-even Higgs bosons, $h$ and $H$, where $h$ denotes the lighter CP-even Higgs boson, one CP-odd Higgs boson $A_0$, two charged Higgs bosons $H^\pm$, and a neutral and two charged Goldstone bosons, $G_0$ and $G^\pm$.  

We replace the original set of parameters  of the Higgs and gauge sector
\begin{align}
m_{11}^2, \quad m_{22}^2, \quad m_{12}^2, \quad \lambda_1, \quad \lambda_2, \quad \lambda_4, \quad  v_1,  \quad v_2,\quad  g_1,  \quad g_2, \quad \lambda_3, \quad \lambda_5
\end{align}
with $g_1$ and $g_2$ being the $U(1)$ and the $SU(2)$ gauge coupling, respectively, by 
\begin{align}
 t_h, \quad t_H, \quad M_h, \quad M_H,  \quad M_{A_0}, \quad M_{H^+}, \quad M_W, \quad M_Z,\quad  e, \quad \beta, \quad \alpha (\text{or } \lambda_3), \quad \lambda_5
\end{align}
with $t_h$ and $t_H$ being the tadpole parameters. The masses of the CP-even, CP-odd, and charged Higgs bosons are $M_h$, $M_H$, $M_{A_0}$, $M_{H^+}$,  the masses of the $Z$ and the $W$ boson are  $M_W$ and $M_Z$. The electric unit charge is  denoted by $e$. The parameter $\beta$ is defined via the ratio of the two Higgs vacuum expectation values, $\tan \beta = \frac{v_2}{v_1}$.  In our different renormalization schemes \footnote{Further renormalization schemes of the THDM are discussed in \citeres{Santos:1996vt,Kanemura:2004mg,LopezVal:2009qy,Degrande:2014vpa,Krause:2016oke,Denner:2016etu,Krause:2016xku,Denner:2017vms}.}, we use either the quartic coupling $\lambda_3$ or the mixing angle of the CP-even Higgs bosons $\alpha$ as an input.

In all four renormalization schemes, 
\begin{itemize}
\item the Higgs- as well as the gauge-boson masses have been chosen on-shell,
\item the electric charge is defined via the electron--positron--photon vertex $ee\gamma$ in the Thomson limit,
\item the quartic coupling $\lambda_5$ is treated as $\overline{\text{MS}}$ parameter.
\end{itemize}

In the applied four different renormalization schemes, we use two different treatments of the tadpoles:
\begin{itemize}
\item 
The renormalized tadpole parameters $t_\phi^{\text{ren}}$ with $\phi = h, H$ vanish. The corresponding counterterm $\delta t_\phi$ is chosen in such a way that the generic one-loop tadpole contributions are canceled. This means that no explicit  tadpole contributions have to be taken into account throughout the calculation. However, this treatment introduces gauge dependences in the relation between bare parameters
\cite{Krause:2016oke,Denner:2016etu}, and, hence, also in the relation between renormalized parameters and physical predictions.
\item 
Following a procedure proposed by Fleischer and Jegerlehner (FJ)~\cite{Fleischer:1980ub},
the bare tadpole parameters  $t_\phi^{\text{bare}}$ vanish~\footnote{In \citere{Actis:2006ra}, a similar scheme, called $\beta_h$~scheme, was suggested.}. The advantage of this treatment is that gauge dependences  in the relation between bare parameters are avoided, and that, thus, the relation between the renormalized parameters and physical predictions does not suffer from gauge dependences. This treatment also requires that explicit tadpole contributions have to be taken into account, which 
can, however, be performed using the same set-up including tadpole counterterms as in the ``$t_\phi^{\text{ren}} = 0$"-prescription, but taking into account finite contributions occurring due to the different treatment of the tadpoles in the \MSbar\ counterterms of $\alpha$, and $\beta$, see Ref.~\cite{Altenkamp:2017ldc} for a detailed description of the procedure.
\end{itemize}

\begin{table}[t]
\begin{center}
{\renewcommand{\arraystretch}{1.8}\begin{tabular}{|l|c|c|c|c|c|}
\hline
& $\alpha$ & $\lambda_3$ & $\beta$ & $t_\phi^{\text{ren}} = 0 $ & $t_\phi^{\text{bare}} = 0 $ \\ \hline
$\MSbar(\lambda_3)$ scheme &  & \MSbar & \MSbar & x &\\\hline
$\MSbar(\alpha)$ scheme & \MSbar & & \MSbar & x & \\\hline
FJ$(\alpha)$ scheme &  \MSbar & &\MSbar & & x \\\hline
FJ$(\lambda_3)$ scheme&& \MSbar  &\MSbar  & &x\\\hline
\end{tabular}}
\end{center}
\caption{Overview about the differences in the different renormalization schemes. It should be noted that the \MSbar~counterterms depend on the choice of the tadpole scheme, i.e.\ whether the renormalized tadpole parameters~$t_\phi^{\text{ren}}$  or the bare tadpole parameter $t_\phi^{\text{bare}}$ are chosen to vanish ($\phi = h, H$). \label{tab:renschemes} }
\end{table}

The following four different renormalization schemes have been 
applied:~\cite{Altenkamp:2017ldc}
\begin{itemize}
\item
$\MSbar(\lambda_3)$ scheme: \\
In this scheme $\lambda_3$ and $\beta$ are independent parameters and fixed in the 
\MSbar{} scheme, and the renormalized tadpole parameters vanish. The mixing angle $\alpha$ can be calculated from $\lambda_3$ and the other independent parameters using tree-level relations. In this scheme, the relation between independent parameters and predicted observables do not depend on a gauge parameter within the class of $R_\xi$ gauges at NLO, since $\lambda_3$ is a basic coupling in the Higgs potential and thus does not introduce gauge dependences, and since the \MSbar{} renormalization of $\beta$ is gauge-parameter independent in $R_\xi$ gauges at NLO~\cite{Krause:2016oke,Denner:2016etu}.
\item
$\MSbar(\alpha)$ scheme: \\
This scheme coincides with the $\MSbar(\lambda_3)$ scheme except that now $\alpha$ is chosen as independent parameter instead of $\lambda_3$.
As explained above, this scheme suffers from some gauge dependence in the relation between renormalized parameters and predicted observables. Hence, for a meaningful comparison with data, all predictions using this renormalization scheme should be performed in the same gauge. 
We use the 't~Hooft--Feynman gauge.
\item
FJ$(\alpha)$ scheme: \\
In this scheme, $\alpha$ and $\beta$ are independent parameters, and the tadpoles are treated following the gauge-independent FJ
prescription, $t_\phi^{\text{bare}} = 0 $.
Similar schemes are also described in \citeres{Krause:2016oke,Denner:2016etu}, however,  the treatment of $m_{12}^2$ and $\lambda_5$ differs.
\item
FJ$(\lambda_3)$ scheme: \\
In this scheme $\beta$ and $\lambda_3$ are independent parameters, as in the
$\MSbar(\lambda_3)$ scheme, but the bare tadpole parameters are chosen to vanish. 
\end{itemize}
An overview of the differences in the four renormalization schemes is given in Tab.~\ref{tab:renschemes}. More details on the different prescriptions
can be found in \citere{Altenkamp:2017ldc}.

The parameters $\alpha$, $\beta$,
and the Higgs-quartic-coupling parameter $\lambda_5$ depend on a renormalization scale $\mu_\mr{r}$ in all four schemes. The $\mu_\mr{r}$~dependence of $\alpha$, $\beta$, 
and $\lambda_5$ is calculated by solving the  renormalization group equations in the four different renormalization schemes \cite{Altenkamp:2017ldc}.

\section{Summary of the calculation}

In this section, we briefly describe the calculation of the decay of the light, neutral 
CP-even Higgs boson of the THDM into four fermions at NLO. The computer program
\Prophecy~\cite{Bredenstein:2006rh,Bredenstein:2006nk,Bredenstein:2006ha}
provides a ``\textbf{PROP}er description of the \textbf{H}iggs d\textbf{EC}a\textbf{Y} into \textbf{4 F}ermions'' and calculates 
observables for the decay process $\Ph {\to} \PW\PW/\PZ\PZ {\to} 4 f$ at NLO EW+QCD in  the SM. 
With our calculation, 
we have extended \Prophecy\ implementing the corresponding decay in the THDM in such a way that the features of 
\Prophecy\ and its applicability basically remain the same. 
We have performed two independent calculations and implementations.
\begin{itemize}
\item For one calculation, 
we have used a model file generated by {\tt  FeynRules} \cite{Christensen2009}, and for the other one an inhouse model file.
\item The amplitudes for the virtual electroweak corrections have been generated with two different versions of {\tt  FeynArts} \cite{Kublbeck:1990xc,Hahn2001}. 
  
  For the virtual QCD corrections, the SM amplitudes of \citere{Bredenstein:2006rh,Bredenstein:2006ha} could be 
reused, and the THDM diagrams were obtained by a proper rescaling of the Higgs couplings. It should be noted, that while masses of 
final-state fermions including the bottom quark mass were neglected in general, the masses were taken into account in closed fermion loops. Hence, the contribution of  diagrams with a closed fermion loop coupling to the Higgs boson does not vanish. Here, special care had to be taken in the rescaling, since the fermion coupling not only scales differently with respect to the 
Higgs--gauge-boson coupling but also depends on the type of the THDM. 
We have implemented four different types (Type 1, Type 2, "flipped", "lepton-specific") that differ in how the down-type and electron-type fermions couple to the two Higgs doublets. Since the up-type fermions couple always in the same manner in all of the four types of THDM and since the largest contribution originates from the 
top-quark loop while the contribution from the other fermions are small, the differences between the types are negligible.
  
  The tree-level and the real contribution were obtained by rescaling  the  Higgs coupling to gauge bosons by  $\sin(\beta - \alpha)$ in the 
SM result of \citeres{Bredenstein:2006rh,Bredenstein:2006ha}.
\item The   amplitude reduction of the electroweak corrections were performed  with {\tt  FormCalc} \cite{Hahn1999,Hahn2000} in the first calculation and with inhouse 
Mathematica routines in the second calculation.
\item The $W$ and $Z$ resonances were treated in the  complex-mass scheme following the prescription in \citere{Denner:2005fg}.
\item The evaluation of loop integrals  was performed with  the public {\tt  Collier} library~\cite{Denner:2016kdg}.
\item Infrared divergences have been treated applying dipole 
subtraction~\cite{Catani1997,Dittmaier:1999mb,Dittmaier:2008md}.
\end{itemize}
More details about the calculation can be found in \citere{Altenkamp:2017kxk}.

%

\section{Numerical results for the partial decay width for $\Ph\to\PW\PW/\PZ\PZ\to4f$}

In this section, we show some sample results for the partial decay width for $\Ph\to\PW\PW/\PZ\PZ\to4f$ for a scenario (scenario A) inspired by  Ref.~\cite{Haber2015} for the Type I THDM:
\begin{equation}
M_h = 125 \text{ GeV}, \quad M_H = 300 \text{ GeV},\quad M_{A_0} = M_{H^+} = 460 \text{ GeV},\quad \lambda_5 = -1.9, \quad \tan\beta= 2.
\end{equation}
Within our calculation, we choose the central renormalization scale as the average mass  of all scalar degrees of freedom, $\mu_0 = (M_h + M_H + M_{A_0} + 2 M_{H^+})/5$.
 
\begin{figure}
\begin{center}
\begin{tabular}{cc}
\includegraphics[width=0.34\textwidth]{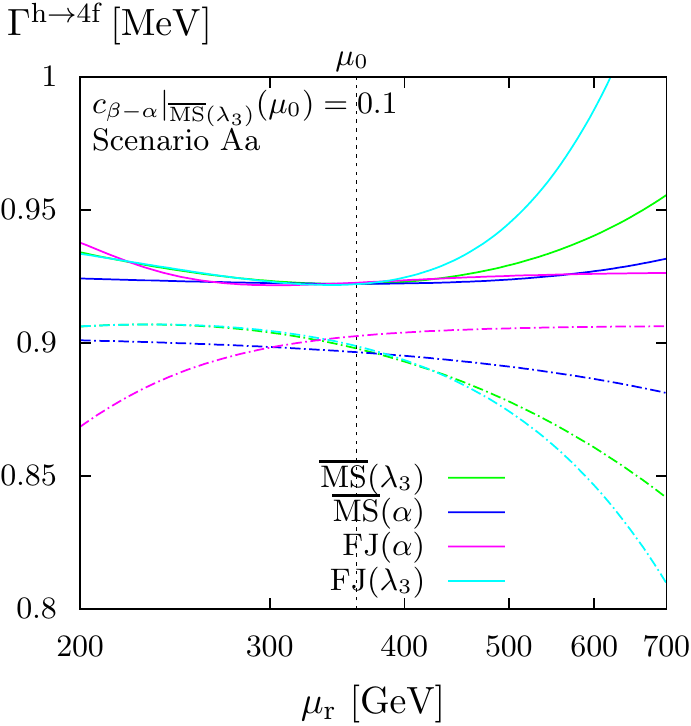} &
\includegraphics[width=0.34\textwidth]{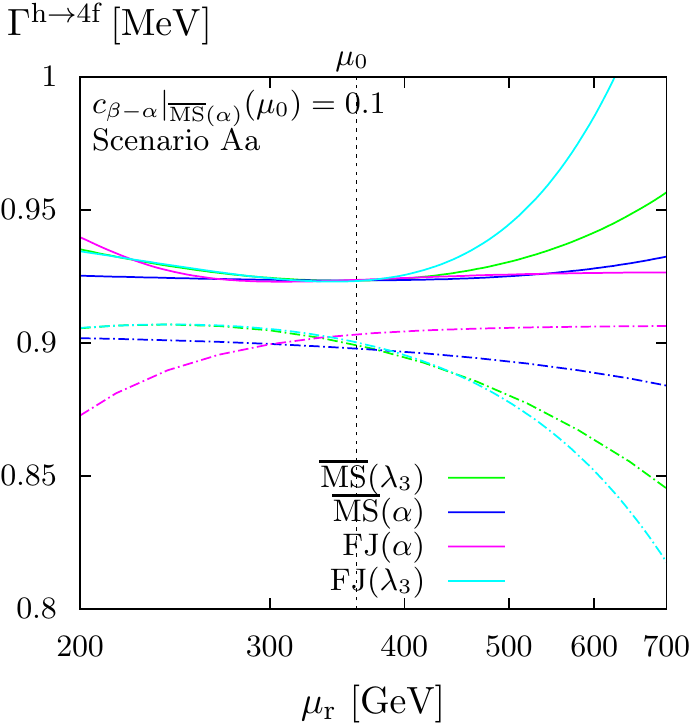} \\
\includegraphics[width=0.34\textwidth]{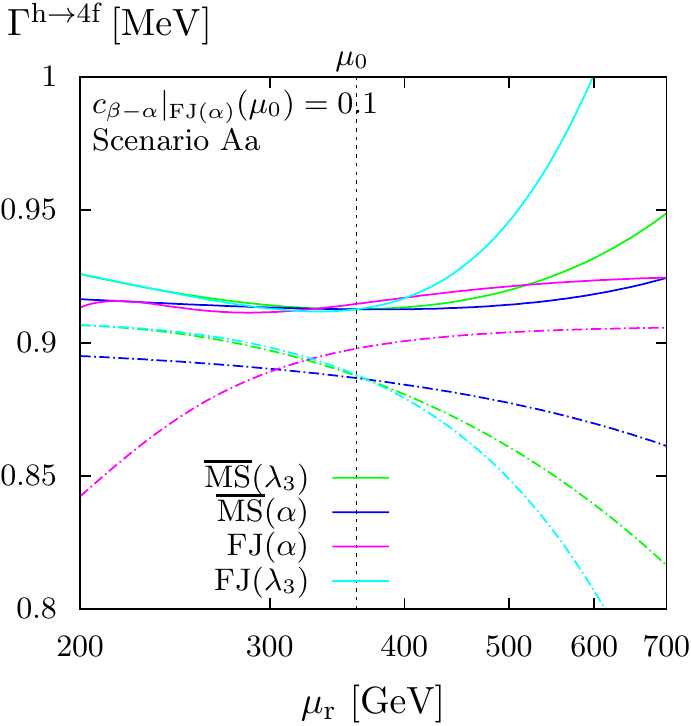}&
\includegraphics[width=0.34\textwidth]{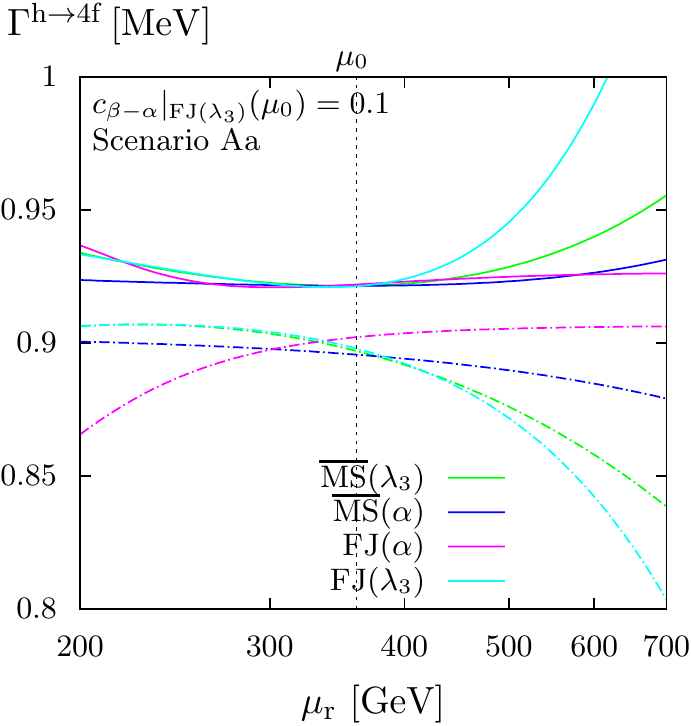}
\end{tabular}
\end{center}
\caption{The renormalization scale dependence of the partial decay width $\Gamma^{\Ph\to4f}_\mr{THDM}$. The four different plots correspond to four different input 
schemes: $\MSbar(\lambda_3)$ (left, top), $\MSbar(\alpha)$ (right, top), FJ$(\alpha)$ (left, bottom), and FJ$(\lambda_3)$ (right bottom). For each plot, the parameters are converted 
to the other schemes,  $\MSbar(\lambda_3)$ (green), $\MSbar(\alpha)$ (blue), FJ$(\alpha)$ (magenta), and FJ$(\lambda_3)$ (turquoise). The solid lines present the result including NLO EW corrections, and the dashed ones show the 
leading-order result. The figure is taken from \protect \citere{Altenkamp:2017kxk}. \label{fig:muscan}}
\end{figure}

In Fig.~\ref{fig:muscan},  the renormalization scale dependence of the partial decay width for $\Ph\to\PW\PW/\PZ\PZ\to4f$, $\Gamma^{\Ph\to4f}_\mr{THDM}$, which is obtained by summing over all partial widths of the h~boson with massless four-fermion final states, is shown. We fix $\cos (\beta - \alpha) = c_{\beta-\alpha} = 0.1$ (scenario Aa). Each plot corresponds to the input parameters given with respect to one of the four renormalization schemes. The dashed curves represent the leading-order (LO) results, however, it should be noted that the input parameters have been converted to the respective scheme denoted by the different line colours. Hence, the strict LO result is only represented by the line corresponding to the input renormalization scheme, i.e.\ for example, in the upper row in the left plot, the strict LO curve is given by the green $\MSbar(\lambda_3)$ line. The differences between the dashed lines at the central renormalization scale $\mu_0$ are only due to conversion effects, while at the other scales also the different running behaviour of the $\MSbar$ parameters in the different schemes plays a role. It should be noted, that it is important to specify not only the parameter values of a certain 
scenario, but also the renormalization scheme, in which these parameters are to be interpreted.

The solid lines show the NLO result including only the EW corrections. A clear plateau around the central renormalization scale $\mu_0$ is 
visible, and there is a clear reduction on the scale dependence going from LO to NLO. 

\begin{figure}\begin{center}
\includegraphics{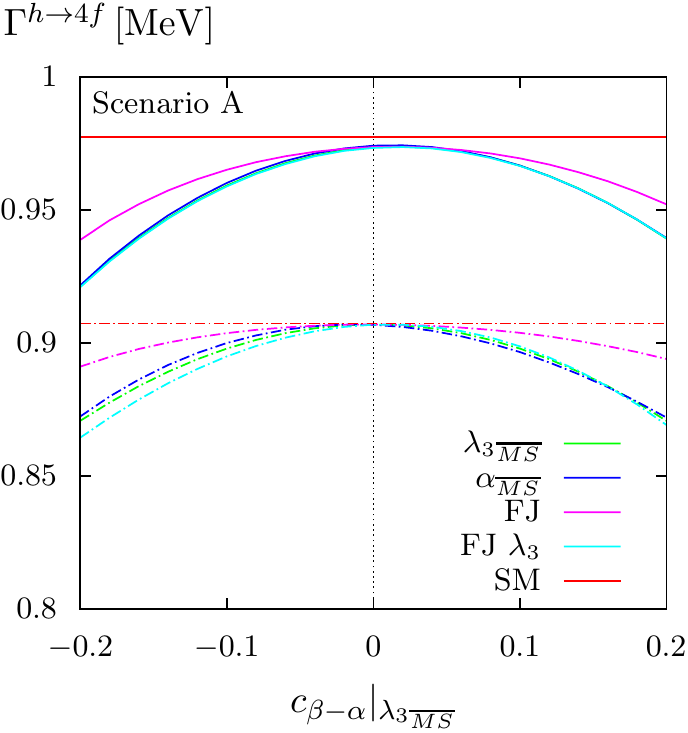}
\end{center}
\caption{The dependence of the partial decay width $\Gamma^{\Ph\to4f}_\mr{THDM}$ on the $\cos(\beta-\alpha)$. The solid (dashed) lines represent the NLO EW + QCD (LO) result. The color code is the same as in Fig.~\ref{fig:muscan}. The SM results are shown in red for comparison. The input parameters are given in the $\MSbar(\lambda_3)$ scheme. The figure is taken from \protect \citeres{Altenkamp:2017ldc,Altenkamp:2017kxk}. \label{fig:cbascan}}
\end{figure}

The complete NLO result including also QCD corrections is shown in Fig.~\ref{fig:cbascan} for the given sample scenario A. The used input scheme is the $\MSbar(\lambda_3)$ scheme. The LO result in the $\MSbar(\lambda_3)$ scheme (dashed, green) corresponds to the SM LO results scaled by the factor $\sin^2(\beta-\alpha)$ and has a parabolic shape. The deviations of the LO results in the other schemes from the $\MSbar(\lambda_3)$ result are again due to the  conversion of the parameters given in the input scheme to the respective final scheme. At NLO, it is interesting to note that, for all schemes, there is a deviation from the SM value also for 
$\cos(\beta- \alpha) = 0$. This deviation originates from the heavy 
Higgs bosons entering the loop contributions. The overall agreement of the results in the different renormalization schemes is better at NLO than at LO. 

A detailed discussion of further results, including also more delicate THDM scenarios, can be
found in \citeres{Altenkamp:2017ldc,Altenkamp:2017kxk}.
The extended version of \Prophecy,
which covers a SM extension with a singlet scalar as well~\cite{Altenkamp:2018bcs}, 
will be available from its hepforge 
webpage~\footnote{\tt http://prophecy4f.hepforge.org/index.html} soon. 

\section{Conclusions}

We have calculated the partial decay width of 
the light CP-even Higgs boson of the THDM decaying into four fermions, $\Gamma^{\Ph\to4f}_\mr{THDM}$, and extended the computer program \Prophecy\ accordingly. 
We have implemented four different renormalization schemes,
imposing on-shell renormalization conditions as far as possible and using
\MSbar\ conditions for mixing angles and quartic scalar self-couplings,
and carried out a
consistent parameter conversion from one scheme to another. In addition, we took into account the running of the $\MSbar$ parameters. The effects of the running 
and of the conversion of parameters can be sizeable depending on the considered scenario. Some sample scenarios have been shown. The overall deviations from the prediction of the SM can be estimated by 
$0$ to $-6\%$ for most of the phenomenologically relevant scenarios. Hereby,  NLO corrections contribute to a shift of one to two percentage points. The size of these deviations is therefore in a range where a linear collider could help in resolving the differences between the models.

\section*{Acknowledgement}
HR would like to thank the organizers of the International Workshop on Future Linear Colliders (LCWS2017) for an interesting and enjoyable workshop.

\section*{References}
\bibliographystyle{h-physrev}            
\bibliography{bibliography}

\begin{thebibliography}{10}

\bibitem{Aad:2015zhl}
ATLAS, CMS, G.~Aad {\em et~al.},
\newblock Phys. Rev. Lett. {\bf 114}, 191803 (2015), 1503.07589.

\bibitem{HHGuide:1990}
J.~F. Gunion, H.~E. Haber, G.~L. Kane, and S.~Dawson,
\newblock {\em {The Higgs Hunter's Guide}} (Westview, 1900).

\bibitem{Gunion:2005ja}
J.~F. Gunion and H.~E. Haber,
\newblock Phys. Rev. {\bf D72}, 095002 (2005), hep-ph/0506227.

\bibitem{Santos:1996vt}
R.~Santos and A.~Barroso,
\newblock Phys. Rev. {\bf D56}, 5366 (1997), hep-ph/9701257.

\bibitem{Kanemura:2004mg}
S.~Kanemura, Y.~Okada, E.~Senaha, and C.~P. Yuan,
\newblock Phys. Rev. {\bf D70}, 115002 (2004), hep-ph/0408364.

\bibitem{LopezVal:2009qy}
D.~Lopez-Val and J.~Sola,
\newblock Phys. Rev. {\bf D81}, 033003 (2010), 0908.2898.

\bibitem{Degrande:2014vpa}
C.~Degrande,
\newblock Comput. Phys. Commun. {\bf 197}, 239 (2015), 1406.3030.

\bibitem{Krause:2016oke}
{Krause, Marcel and Lorenz, Robin and M\"uhlleitner, Margarete and Santos, Rui
  and Ziesche, Hanna},
\newblock JHEP {\bf 09}, 143 (2016), 1605.04853.

\bibitem{Denner:2016etu}
A.~Denner, L.~Jenniches, J.-N. Lang, and C.~Sturm,
\newblock JHEP {\bf 09}, 115 (2016), 1607.07352.

\bibitem{Krause:2016xku}
{Krause, Marcel and M\"uhlleitner, Margarete and Santos, Rui and Ziesche,
  Hanna},
\newblock Phys. Rev. {\bf D95}, 075019 (2017), 1609.04185.

\bibitem{Denner:2017vms}
A.~Denner, J.-N. Lang, and S.~Uccirati,
\newblock JHEP {\bf 07}, 087 (2017), 1705.06053.

\bibitem{Fleischer:1980ub}
J.~Fleischer and F.~Jegerlehner,
\newblock Phys. Rev. {\bf D23}, 2001 (1981).

\bibitem{Actis:2006ra}
S.~Actis, A.~Ferroglia, M.~Passera, and G.~Passarino,
\newblock Nucl. Phys. {\bf B777}, 1 (2007), hep-ph/0612122.

\bibitem{Altenkamp:2017ldc}
L.~Altenkamp, S.~Dittmaier, and H.~Rzehak,
\newblock JHEP {\bf 09}, 134 (2017), 1704.02645.

\bibitem{Bredenstein:2006rh}
A.~Bredenstein, A.~Denner, S.~Dittmaier, and M.~M. Weber,
\newblock Phys. Rev. {\bf D74}, 013004 (2006), hep-ph/0604011.

\bibitem{Bredenstein:2006nk}
A.~Bredenstein, A.~Denner, S.~Dittmaier, and M.~M. Weber,
\newblock Nucl. Phys. Proc. Suppl. {\bf 160}, 131 (2006), hep-ph/0607060.

\bibitem{Bredenstein:2006ha}
A.~Bredenstein, A.~Denner, S.~Dittmaier, and M.~M. Weber,
\newblock JHEP {\bf 02}, 080 (2007), hep-ph/0611234.

\bibitem{Christensen2009}
N.~D. Christensen and C.~Duhr,
\newblock Comput.Phys.Commun. {\bf 180}, 1614 (2009), 0806.4194.

\bibitem{Kublbeck:1990xc}
J.~{K\"ublbeck}, M.~{B\"ohm}, and A.~Denner,
\newblock Comput. Phys. Commun. {\bf 60}, 165 (1990).

\bibitem{Hahn2001}
T.~Hahn,
\newblock Comput. Phys. Commun. {\bf 140}, 418 (2001), hep-ph/0012260.

\bibitem{Hahn1999}
T.~Hahn and M.~Perez-Victoria,
\newblock Comput. Phys. Commun. {\bf 118}, 153 (1999), hep-ph/9807565.

\bibitem{Hahn2000}
T.~Hahn,
\newblock Nucl. Phys. Proc. Suppl. {\bf 89}, 231 (2000), hep-ph/0005029.

\bibitem{Denner:2005fg}
A.~Denner, S.~Dittmaier, M.~Roth, and L.~H. Wieders,
\newblock Nucl. Phys. {\bf B724}, 247 (2005), hep-ph/0505042,
\newblock [Erratum: Nucl. Phys.B854,504(2012)].

\bibitem{Denner:2016kdg}
A.~Denner, S.~Dittmaier, and L.~Hofer,
\newblock Comput. Phys. Commun. {\bf 212}, 220 (2017), 1604.06792.

\bibitem{Catani1997}
S.~Catani and M.~H. Seymour,
\newblock Nucl. Phys. {\bf B485}, 291 (1997), hep-ph/9605323.

\bibitem{Dittmaier:1999mb}
S.~Dittmaier,
\newblock Nucl. Phys. {\bf B565}, 69 (2000), hep-ph/9904440.

\bibitem{Dittmaier:2008md}
S.~Dittmaier, A.~Kabelschacht, and T.~Kasprzik,
\newblock Nucl. Phys. {\bf B800}, 146 (2008), 0802.1405.

\bibitem{Altenkamp:2017kxk}
L.~Altenkamp, S.~Dittmaier, and H.~Rzehak,
\newblock (2017), 1710.07598.

\bibitem{Haber2015}
H.~E. Haber and O.~St{\aa}l,
\newblock Eur. Phys. J. {\bf C75}, 491 (2015), 1507.04281.

\bibitem{Altenkamp:2018bcs}
L.~Altenkamp, M.~Boggia, and S.~Dittmaier,
\newblock (2018), 1801.07291.

\end{thebibliography}

\end{document}